# Specific Heat vs Field in the 30 K Superconductor $BaFe_2(As_{0.7}P_{0.3})_2$

J. S. Kim[1], P. J. Hirschfeld[1], G. R. Stewart[1], S. Kasahara[2], T. Shibauchi[3], T. Terashima[2], and Y. Matsuda[3]

[1]Department of Physics, University of Florida
Gainesville, FL 32611-8440
[2] Research Center for Low Temperature and Materials Sciences,
Kyoto University, Sakyo-ku, Kyoto 606-8501, Japan
[3] Department of Physics, Kyoto University,
Sakyo-ku, Kyoto 606-8502, Japan

**Abstract:** We report specific heat measurements on the Fe-based superconductor $BaFe_2(As_{0.7}P_{0.3})_2$, a material on which previous penetration depth, NMR, and thermal conductivity measurements have observed a high density of low-energy excitations, which have been interpreted in terms of order parameter nodes. Within the resolution of our measurements, the low temperature limiting C/T is found to be linear in field, i.e. we find no evidence for a Volovik effect associated with nodal quasiparticles in either the clean or dirty limit. We discuss possible reasons for this apparent contradiction.

## I. Introduction

The investigation of a new class of superconductors, such as the recently discovered[1] 'high' temperature superconductors based on iron pnictides (FePn), inevitably leads to concerted efforts to establish the symmetry of the order parameter. In the case of the FePn materials, these investigations are still returning apparently conflicting results, even in samples that are essentially identical. For example, Andreev spectroscopy[2] implies a fully gapped superconductor in the 1111 material $SmFeAsO_{0.85}F_{0.15}$, whereas at almost the same composition ($SmFeAsO_{0.82}F_{0.18}$) infrared optical measurements[3] imply a nodal superconductor. London penetration depth measurements[4] in the 122 structure $Ba(Fe_{0.93}Co_{0.07})_2As_2$ suggest nodes in the superconducting gap, while thermal conductivity measurements[5] on samples that include the same composition ($Ba(Fe_{1-x}Co_x)_2As_2$, $0.048 \leq x \leq 0.114$) from the same group as ref. 4 imply a fully gapped superconductor. There are many other examples of apparently conflicting results in both the 1111[6-10] and the 122[11-19] structures where the differing conclusions may be a question of sample doping, quality, or other variation. Another possibility, however, is that different measurements probe different *parts* of the Fermi surface.

Recently, superconductivity at 30 K has been reported[20] in P-doped $BaFe_2As_2$, with 1/3 of the As replaced by P. This system is particularly interesting because it displays a phase diagram similar to that of other doped ferropnictides, but the process by which the system is "doped" is far from obvious, given that P and As are isoelectronic. In addition, transport measurements[20] show indications of quantum critical behavior (consistent with dHvA measurements[21]) near the maximum critical temperature, similar to hole-doped cuprate superconductors. Magnetic penetration depth and thermal conductivity measurements[22], as well as NMR data[23], indicate superconducting gaps with nodes. Here we report measurements of the specific heat divided by temperature, C/T, as a function of field in order to further investigate the gap structure with a bulk probe which is sensitive to all electronic excitations. Specific heat in field was pioneered[24] as a tool for investigating the gap structure on YBCO, with the theory of Volovik[25] predicting $\gamma$ (=C/T as T→0) ~ $H^{1/2}$ in a clean superconductor with lines of nodes, while the theory of Kübert and Hirschfeld[26] gives $\gamma$ ~ HlogH for a disordered superconductor with lines of nodes. These power laws arise from the Doppler shift of the low-energy nodal quasiparticles in the superflow field of the vortex lattice. For a fully gapped superconductor, $\gamma$ will vary simply as H due to the localized Caroli-de Gennes-Matricon states in the vortex cores[27].

## II. Experimental

Tiny platelet crystals of $BaFe_2(As_{0.67}P_{0.33})_2$ were prepared as in ref. 20. A collage of 21 mg of these microcrystals was mounted on a sapphire disk using GE7031 varnish. Approximately 75% of the crystals had the magnetic field perpendicular to the a-b plane (the plane of the crystals), with the rest randomly oriented. The sapphire disk was mounted in our time constant method calorimeter[28], and the specific heat from 0.4 to 5 K in fields from 0 to 15 T was measured. Additionally, the specific heat of a standard (high purity Au) was measured in fields up to 14 T. Results on the standard (not shown) indicate agreement with published values to within ±3% in all fields.

## III. Results and Discussion

The specific heat of $BaFe_2(As_{0.7}P_{0.3})_2$ for 0 ≤ H ≤ 15 T is shown in Fig. 1.

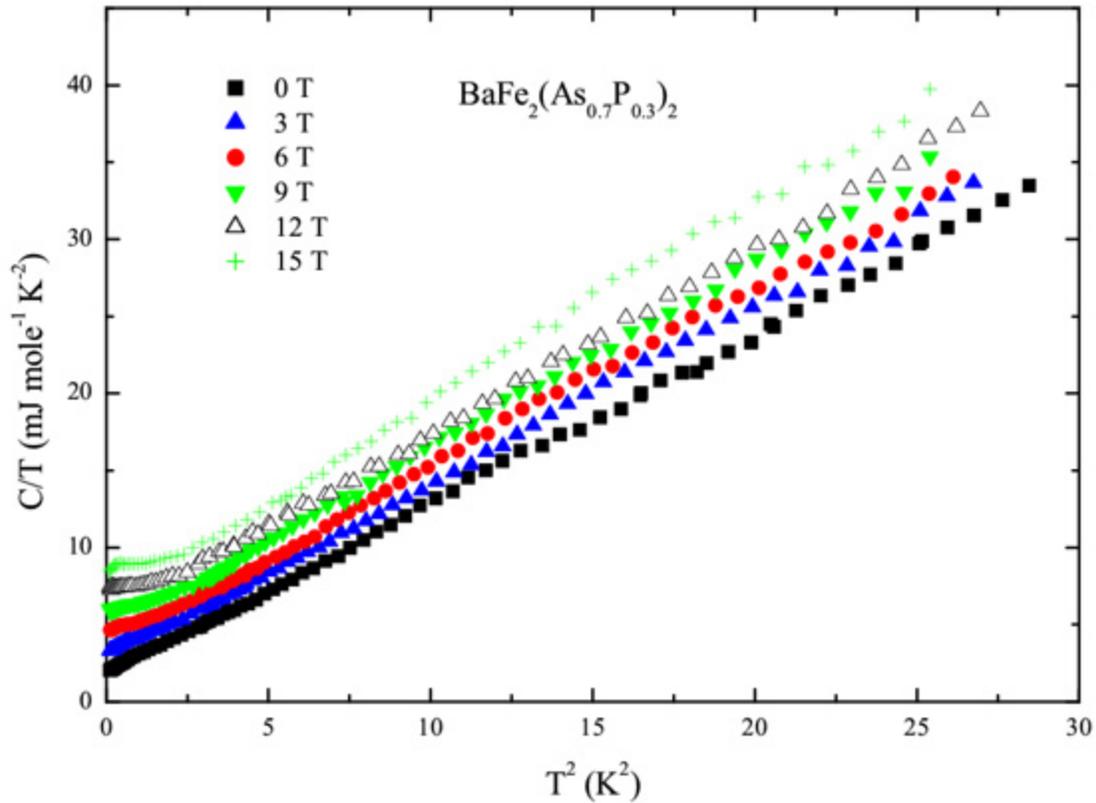

Fig. 1(color online): Specific heat divided by temperature vs the square of the temperature of BaFe$_2$(As$_{0.7}$P$_{0.3}$)$_2$ as a function of field in 3 T steps. Data at another five fields are not shown for clarity.

As will be discussed further below, there is a small low temperature anomaly in the specific heat data below about 1.4 K. Low temperature anomalies have been seen[29] in other FePn samples, and in some cases – for example in BaFe$_{2-x}$Co$_x$As$_2$ – show rather strong magnetic field dependence.[29] In order to obtain an idea what the temperature dependence of the electronic specific heat is in 0 field, and to accentuate the small, T<1.4 K, anomaly, Fig. 2 shows a log-log plot of the data corrected for a slight residual linear term $\gamma T$ ($\gamma$=1.78 mJ/moleK$^2$) in the superconducting specific heat and the approximate lattice contribution to the specific heat. This plot shows clearly the low

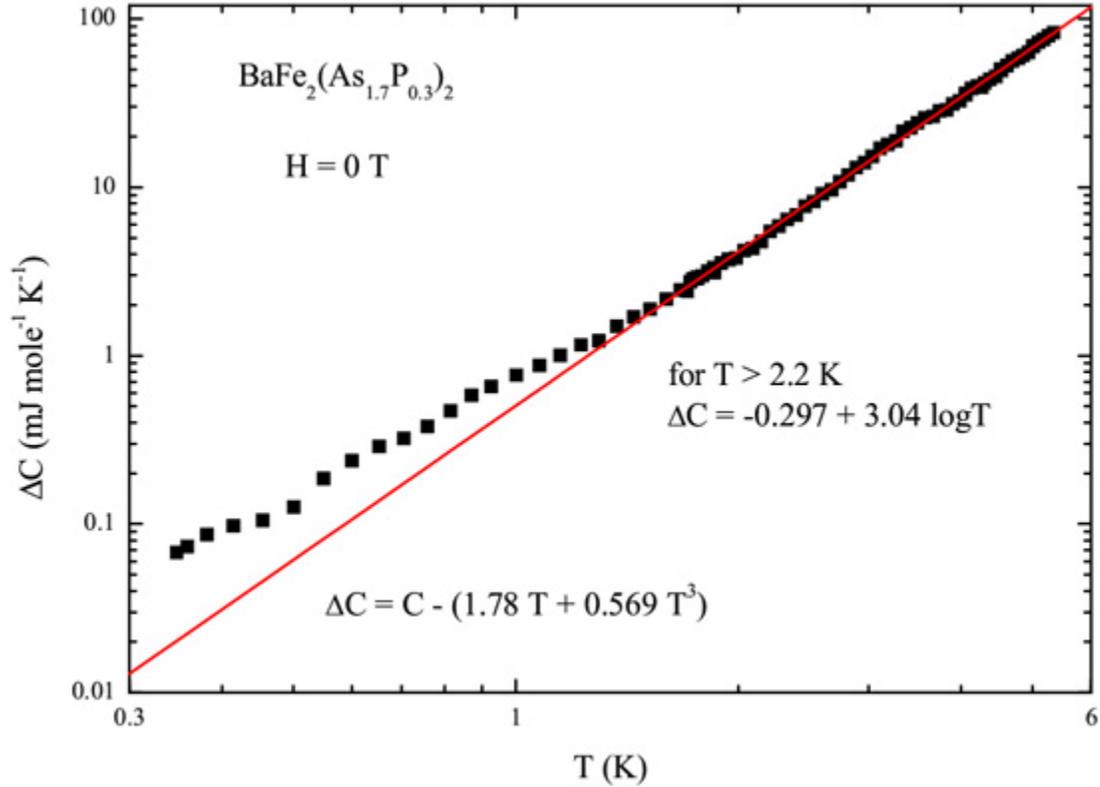

Fig. 2 (color online). Log of $\Delta C$ (= $C_{measured}$ - $\gamma T$ – $C_{lattice}$) vs logT, where $\gamma$ is obtained by extrapolating C/T from the lowest temperature data to T=0 and the lattice specific heat is approximated by using a Debye temperature of 250 K for $BaNi_2As_2$[30] and scaling it for $BaFe_2(As_{0.7}P_{0.3})_2$ by the square root of the inverse ratio of the molar masses to arrive at ~ 257 K, or $C_{lattice} = \beta T^3$, with $\beta$=0.569 mJ/molK$^4$. Note that the small anomaly in the specific heat appears to be finished by about 1.4 K.

temperature anomaly in the zero field data. In any case, the difference $\Delta C$, the superconducting electronic specific heat, is quite close to $T^3$ in temperature dependence. Attempts to separate out a $T^2$ contribution, which was very difficult to achieve[31] in the high $T_c$ cuprate superconductors until superior samples were grown, were unsuccessful here.

In order to more closely consider the field dependence of the specific heat in Fig. 1, these data are shown in Fig. 3 with the zero field data subtracted. The data in Fig. 3

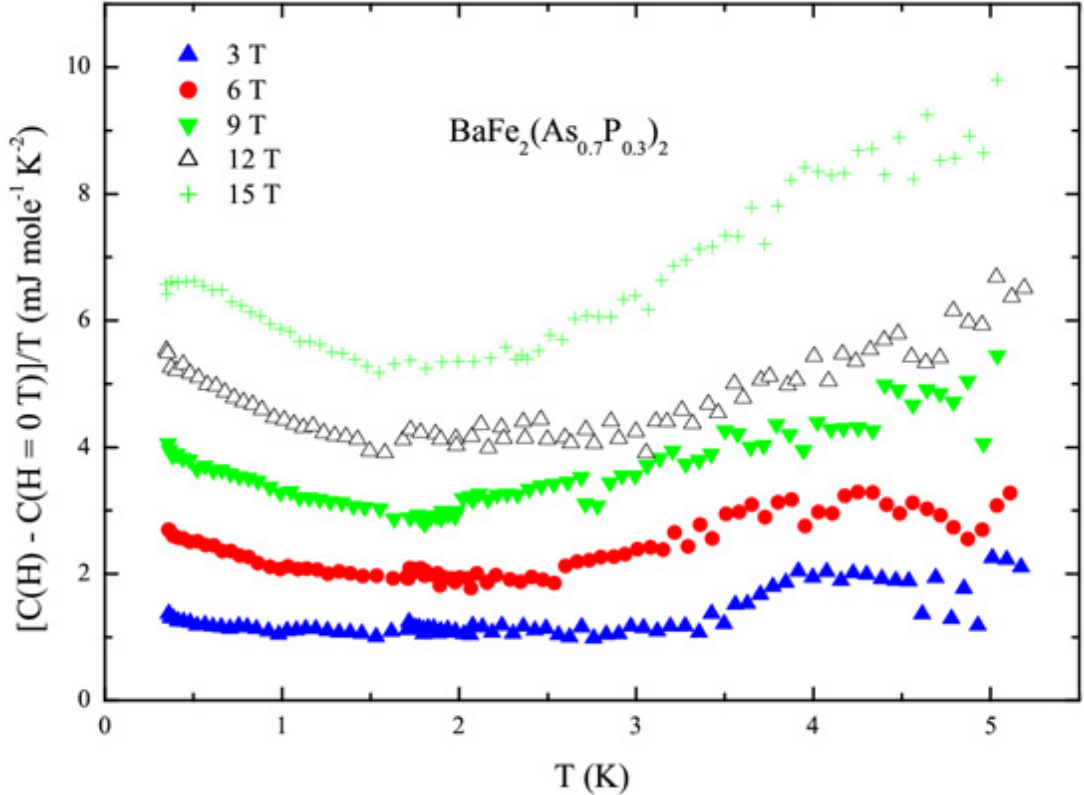

Fig. 3 (color online): Specific heat divided by temperature vs temperature, with the zero field result subtracted. Note the increase of C/T at low temperatures with increasing field (the field dependence of which is the goal of the measurements of this work and discussed below in Fig. 4.) The low temperature anomaly appears to be almost field independent since the upturn in the data shown here below 1 K which increases with increasing field is due to the specific heat contribution from the splitting of the nuclear levels with field ($C_{nuc}/T \sim H^2/T^3$), which is calculable (3.7 mJ/moleK$^2$ at 0.4 K and 15 T) for the isotopes involved ($^{135}$Ba, 6.6% abundant, $^{137}$Ba, 11.2 % abundant, $^{57}$Fe, 2.14 % abundant, $^{75}$As, 100% abundant, and $^{31}$P, 100% abundant).

show that the low temperature anomaly in C, as well as the nuclear hyperfine splitting due to the magnetic field, appear to have little influence around 1.5 to 2 K. Thus, in order to track the behavior of $\gamma$ ($\equiv$C/T as T$\to$0) we propose three equivalent ways to treat the data. By extrapolating the data (C/T = $\gamma$ + $\beta T^2$ + $\delta T^4$) by a three term fit from 1.5 K and above, we avoid the low temperature anomaly and can well approximate $\gamma$. Another method is to just take the value of C/T at either 1.5 K or 2 K as indicative of γ with no influence from either the anomaly or the field-induced nuclear contribution. Such a plot of γ as a function of field is shown in Fig. 4.

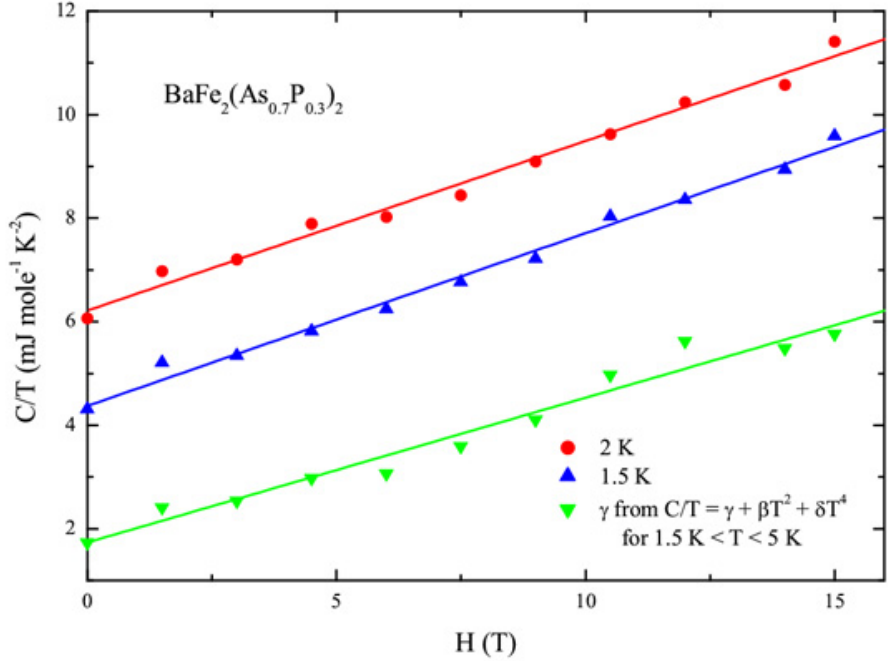

Fig. 4(color online): Three methods of determining the behavior of γ in BaFe$_2$(As$_{0.67}$P$_{0.33}$)$_2$ as a function of field up to 15 T. All methods result in γ ~ H. Extrapolating the specific heat T=0 γ (upside down triangles) to the upper critical field H$_{c2}$ of about 52 T (from ref. 22) gives a value of 16 mJ/moleK$^2$ for the normal state γ, which is comparable to the value of 18 mJ/moleK$^2$ determined[17] recently for Ba(Fe$_{0.92}$Co$_{0.08}$)$_2$As$_2$.

IV. Analysis

Here we discuss the extent to which the results presented here are consistent with those found in thermal conductivity, NMR, and penetration depth measurements on the same samples. In each of these three cases, power laws in temperature were reported at low temperatures $T<<T_c$, consistent with line nodes somewhere on the Fermi surface. In the case of the NMR $(T_1T)^{-1}$ measurements, the behavior at the lowest temperatures indicated further the presence of substantial amounts of disorder, consistent with an impurity bandwidth of several Kelvin, whereas the linear-$T$ behavior in the penetration depth, characteristic of a clean nodal system, appears to extend down to about 1K before deviations characteristic of an impurity band are visible. The linear-T term at the lowest temperatures in the thermal conductivity appears to be substantially larger than the universal value expected in $d$-wave superconductors, but could be consistent with the nonuniversal result found for extended-$s$ states obtained in Ref. 32. In the present work, a substantial residual $\gamma$ (1.78 mJ/moleK$^2$) value has been obtained, of order similar to that observed in cuprate samples. This is again consistent with the presence of line nodes in the superconducting order parameter, with a temperature range of order a few Kelvin where quasiparticle states are broadened by disorder. While power laws observed in low-temperature NMR and thermal conductivity might possibly be consistent with a disorder dominated isotropic state, the linear-T term reported in the penetration depth experiment[22] requires a state with true line nodes.

On the other hand, the linear field dependence measured here is not obviously consistent with such a picture. In a fully gapped superconductor, the electronic specific heat at low temperatures $T<<T_c$ is proportional to the applied field, since quasiparticles are confined to the vortex cores where the order parameter is suppressed, and flux quantization requires that the field scales with the number of vortices. Approximating the density of states in the core by that of the normal metal above $T_c$, one finds $C(T)/T \sim N_0 (H/H_{c2})$, where $N_0$ is the the density of states at the Fermi surface and $H_{c2}$ is the upper critical field. Thus the power law in field measured here might, in isolation, be taken as evidence for a fully developed spectral gap. In a superconductor with gap nodes, the

Doppler shift of the energies of quasiparticles outside the vortex core moving in the superflow field of the vortex lattice leads to a a stronger field dependence, $C(T)/T \sim N_0 (H/H_{c2})^{1/2}$. A similar effect occurs in the thermal conductivity of a nodal superconductor, and was in fact observed clearly in Ref. 22 on samples similar to those measured here. Disorder can alter the result for the electronic specific heat to $\sim H \log H$ over a field scale of order the impurity bandwidth, but should still impart a substantial downward curvature to $C/T$ vs. $H$ with increasing field. The fact that this is not observed here is therefore at first glance inconsistent with all other measurements on $BaFe_2(As_{0.7}P_{0.3})_2$.

A possible resolution to this discrepancy may be found by recognizing that the excitations which contribute to the electronic specific heat arise from all parts of the Fermi surface. Many spin fluctuation theories of the Fe-based superconductors find that the hole pockets around the $\Gamma$ point are fully gapped, whereas the electron pockets around M support anisotropic gaps, possibly with nodes. Thus in a simple model, the specific heat of this system should consist of two terms, $C(T)=C_\alpha(T)+C_\beta(T)$, where $\alpha$ refers to the hole sheets and $\beta$ to the electron sheets, and we might expect $C_\alpha/T \sim H$ and $C_\beta/T \sim \sqrt{H}$. Thus it is possible that if the masses on the respective sheets are very different, such that $N_0^\alpha \gg N_0^\beta$, the specific heat will be dominated by the hole sheets, and yield a result apparently characteristic of a fully gapped state. In such a scenario, the finite specific heat $\gamma$ observed in the zero field measurements would have to be ascribed to extrinsic sources, such as two-level systems.

This simple light electron (nodal)/heavy hole (fully gapped) sheet model would be consistent with the observation of T dependence in NMR spin-lattice relaxation time measurements[23] characteristic of nodes, since the contribution from the heavy fully gapped hole sheet would be exponentially suppressed at low T. In addition, the clear nonlinear field dependence of the thermal conductivity[22] can be understood, since quasiparticle transport currents (e.g. low-temperature thermal currents) will be dominated by the lighter mass electron sheets. Longer quasiparticle lifetimes on the electron sheets may also play a role in this case.[33] Recent dHvA experiments[34] have succeeded in measuring the mass of one of the hole sheets at an overdoped concentration ($BaFe_2(As_{0.4}P_{0.6})_2$), and find that it has approximately twice the effective mass as the electron sheets. Furthermore, an additional hole sheet expected from density functional theory has not yet been observed, possibly because its mass is too high. Clearly, further studies are necessary to substantiate this point of view.

### V. Conclusions

Specific heat of $BaFe_2(As_{0.67}P_{0.33})_2$ in fields to 15 T reveal that, in contrast to the penetration depth and thermal conductivity measurements, the part of the Fermi surface sampled by the specific heat reveals mixed evidence of nodal superconductivity. While the large residual $\gamma$ in the superconducting state could be due to a disordered superconducting state with nodes, we observe a linear field dependence of the specific heat at low temperatures, and no evidence for the Volovik effect characteristic of nodal quasiparticles which should lead to negative curvature in C(H). We discussed a scenario in which the field dependence in our measurement might be dominated by heavy mass hole sheets of the Fermi surface corresponding to an isotropic order parameter, while electron sheets with a highly anisotropic order parameter could be difficult to observe due to a much lighter mass. Such an attempt to reconcile existing measurements on these samples requires more detailed studies probing the character of the states on the various parts of the Fermi surface in these materials.

**Acknowledgements:** Work at University of Florida by JSK and GRS supported by the US Department of Energy, contract no. DE-FG02-86ER45268. PJH was supported by DE--FG02-05ER46236.